\documentclass[twocolumn,aps,showpacs,prb,tightenlines,amsmath,amssymb]{revtex4}
\usepackage{graphicx}
\usepackage{amssymb}
\usepackage{dcolumn}
\usepackage{amsmath}
\usepackage{bm}

\usepackage{colordvi}

\begin{document}

\title{Effect of Singwi-Tosi-Land-Sj\"{o}lander local field correction on
spin relaxation in $n$-type GaAs quantum wells at low temperature}
\author{J. Zhou}
\affiliation{Hefei National Laboratory for Physical Sciences at
  Microscale and Department of Physics, University of Science and Technology of China,
Hefei, Anhui, 230026, China}
\date{\today}

\begin{abstract}
We study the effect of the Singwi-Tosi-Land-Sj\"{o}lander local field correction on
  spin relaxation/dephasing in $n$-type GaAs quantum wells at low temperature by
  constructing and numerically solving the kinetic spin Bloch
  equations.
  We calculate the local field factor $G(q)$ in quantum wells
by numerically solving three equations which link the
 local field factor, the structure factor, and
 the dielectric function, self-consistently.
Such a correction reduces
 both the electron-electron Coulomb scattering and the Coulomb Hartree-Fock
term. We compare the spin
 relaxation times with and without this
  correction under different conditions such as temperature, electron
  density, well width and spin polarization. We find that this correction
  leads to a decrease/increase of the spin relaxation time in the strong/weak scattering
  limit. At high spin polarization, it reduces the Hartree-Fock term and
consequently tends to decrease the spin relaxation time.
The modification of the
spin relaxation time by the local field correction is more or less
 moderate either due to the coexistence of scattering other
than the Coulomb scattering at low spin
polarization and/or due to the competing effects
from the Coulomb scattering and
the Coulomb Hartree-Fock term at high polarization.

\end{abstract}


\maketitle

The study of spin relaxation/dephasing (R/D) in semiconductors is of great
interest to the spintronic society recently.\cite{meier,prinz,Fabian}
It is well known that the
D'yakonov-Perel' (DP) mechanism\cite{DP} is the dominant spin
R/D mechanism in both bulk and two-dimensional
$n$-type zinc-blende semiconductors.\cite{zhou}
This mechanism is due to the momentum-dependent spin state splitting,
which originates from the Dresselhaus spin-orbit coupling\cite{dress}
in crystals without inverse symmetry and the Rashba
spin-orbit coupling\cite{rashba} in quantum wells (QWs) with
asymmetric potential.
For (001) GaAs QW with small well width, the Dresselhaus term can be
written as\cite{dp2}
\begin{eqnarray}
\label{omegax}
\Omega_x({\bf k})&=&\gamma k_x(k_y^2-\langle k_z^2\rangle), \\
\Omega_y({\bf k})&=&\gamma k_y(\langle k_z^2\rangle-k_x^2),  \\
\label{omegaz}
\Omega_z({\bf k})&=&0\ ,
\end{eqnarray}
in which $\langle k_z^2\rangle$ represents the average of the operator
$-(\partial/\partial z)^2$ over the electronic state of the lowest
subband.\cite{wu1}

It was first pointed out by Wu {\em et al.} from
a fully microscopic approach that
any type of scattering, including the spin conserving electron-electron Coulomb
scattering, can lead to irreversible spin R/D in the presence of inhomogeneous
broadening.\cite{wu2,wu3,wu4,wu5,wu6} Similar claim was also made later by
Glazov and Ivchenko.\cite{ivch}
This has been verified experimentally by Leyland  {\em et al.}.\cite{harley}
Moreover, Weng and Wu also predicted an interesting
effect that the spin R/D can be
suppressed by increasing the initial spin polarization due to the
effective magnetic field which originates from the Hartree-Fock (HF)
 contribution of the Coulomb interaction.\cite{wu3}
This prediction has also been confirmed experimentally very
recently.\cite{wu8} However, in all the previous
works,\cite{wu1,wu2,wu3,wu4,wu5,wu6,wu8} the electron-electron
Coulomb interaction is treated under the random phase approximation
(RPA). Usually the RPA is considered to be good only at small $r_s$,\cite{comment} with $r_s=(\sqrt{\pi n}a_{\mbox
    {\tiny B}})^{-1}$ representing the dimensionless coupling parameter for
the two-dimensional electron gas with electron density
  $n$ and effective Bohr radius $a_{\mbox{\tiny B}}$.
Such kind of approximation does not involve
the correction which describes the electron charge depletion nearby
the electron, known as the exchange-correlation hole.
This correction beyond the RPA is so-called the local field correction (LFC).
It must be taken into account for
properties when the electron-electron interactions plays an important role,
{\em eg.}, the spin Coulomb drag effect.\cite{badalyan}
This LFC can be introduced through a
momentum-dependent correction factor $G(q)$ in the dielectric
function:\cite{hubbard,sham,geldart,toigo,STLS,STLS1,vashista,utsumi}
$\varepsilon({\bf q},\omega)=1-{v_{q}P^{(1)}({\bf q},\omega)}/{[1
+v_{q}G(q)P^{(1)}({\bf q},\omega)]}$,
where $P^{(1)}({\bf q},\omega)$ is the free-electron dynamical
polarizability and $v_{q}=4\pi e^2/q^2$ is the bare Coulomb potential.
This  factor was originally introduced by Hubbard\cite{hubbard}
in very simple form
$G_{\mbox{\tiny H}}(q)=\frac{1}{2}\frac{q^2}{q^2+k_{F}^{2}}$ with
$k_{F}$ being the Fermi wave vector.
This form is unappreciated nowadays due to the lack of accuracy.
 Later, several improved forms of
$G(q)$ based on the diagrammatic techniques were
proposed.\cite{sham,geldart}
Singwi {\em et al.}\cite{STLS} gave a different and more
precise choice of local field factor $G_{\mbox{\tiny STLS}}(q)$
as a functional of static structure factor $S(q)$
by using the equation of motion method:
\begin{equation}
G_{\mbox{\tiny STLS}}(q)=\frac{-1}{n}\int \frac{d{\bf q^{\prime}}}
{(2\pi)^3}\frac{{\bf
    q}\cdot{\bf q^{\prime}}}{q^{\prime 2}}[S({\bf q}-{\bf q^{\prime}})-1]\ .
\label{eqSTLS}
\end{equation}
They suggested that one can solve a triad
of equations which link the three quantities: $G_{\mbox{\tiny
    STLS}}(q)$, the structure factor $S(q)$ and
$\varepsilon({\bf q},\omega)$ self-consistently.
Later on, there are  many other efforts on calculating this factor
in three-dimensional (3D)
systems.\cite{vashista,utsumi,oritz,bretonnet,hellal,moroni,wardana}
In ideal two-dimensional (2D) electron gas/liquid, the local field
factors $G_{\mbox{\tiny 2D}}(q)$ were also
calculated\cite{bulutay,davoudi,yur,atwal,moreno,asgari,wardana1}
similar to the 3D case mentioned above.

In this paper, we calculate $G^{\mbox{\tiny W}}(q)$ in $n$-type GaAs
QWs by solving three equations which link the three functions
$G^{\mbox{\tiny W}}(q)$, $S^{\mbox{\tiny W}}(q)$, and
$\varepsilon^{\mbox{\tiny W}}({\bf q},\omega)$ self-consistently.
These equations which are a little different from those in the
bulk system\cite{STLS} and ideal 2D system\cite{bulutay} can be written
as:
\begin{eqnarray}
&&\varepsilon^{\mbox{\tiny W}}({\bf q},\omega)=1-\frac{\bar{v}_q\bar{P}^{(1)}({\bf
    q},\omega)}{1+\bar{v}_qG^{\mbox{\tiny W}}(q)\bar{P}^{(1)}({\bf
    q},\omega)}\ , \\
&&G^{\mbox{\tiny W}}(q)=\frac{-1}{n}\int \frac{d{\bf
    q}^{\prime}}{(2\pi)^2}\frac{{\bf q} \cdot {\bf q}^{\prime}\bar{v}_{q^{\prime}}}{q^{
  2}\bar{v}_q}\big [S^{\mbox{\tiny W}}({\bf q}-{\bf q^{\prime}})-1 \big ]\ ,\\
&&S^{\mbox{\tiny W}}({\bf q})=\frac{-1}{n\bar{v}_q}\int_{0}^{\infty}\frac{d\omega}{\pi}
\mbox{Im}\big [\frac{1}{\varepsilon^{\mbox{\tiny W}}({\bf q},\omega)}\big]\ ,
\end{eqnarray}
where $\bar{v}_q=\sum_{q_z}v_{Q}|I(iq_z)|^2$ and $v_{Q}=4\pi e^2/Q^2$
with ${\bf Q}\equiv ({\bf q}, q_z)$ are the screened quasi-2D and bare
bulk Coulomb potential respectively.
$|I(iq_z)|^2$ denotes the form factor of square QWs with
finite well depth, whose
expression can be found in Ref.\ [\onlinecite{wu1}].

\begin{figure}[htb]
\includegraphics[height=5.5cm]{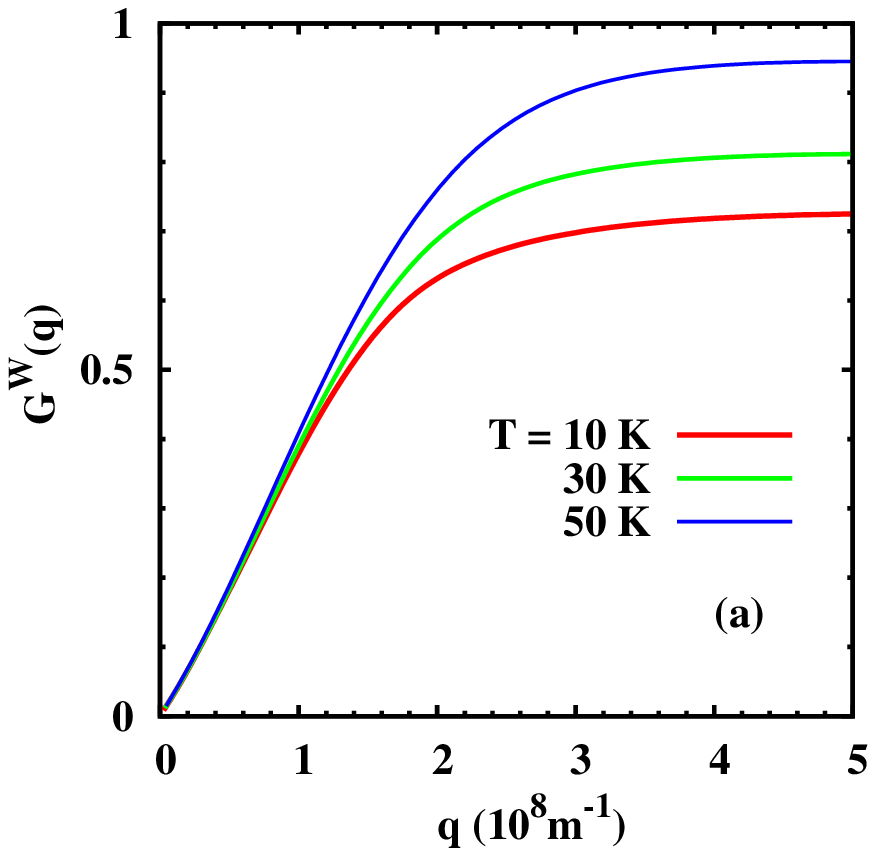}
\includegraphics[height=5.5cm]{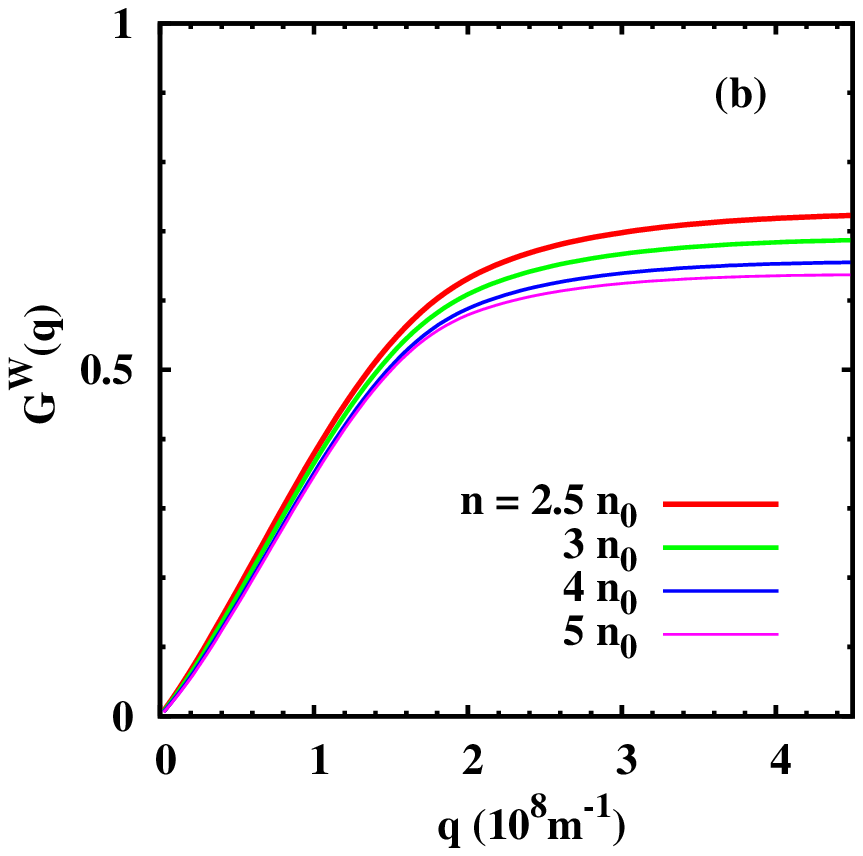}
\includegraphics[height=5.5cm]{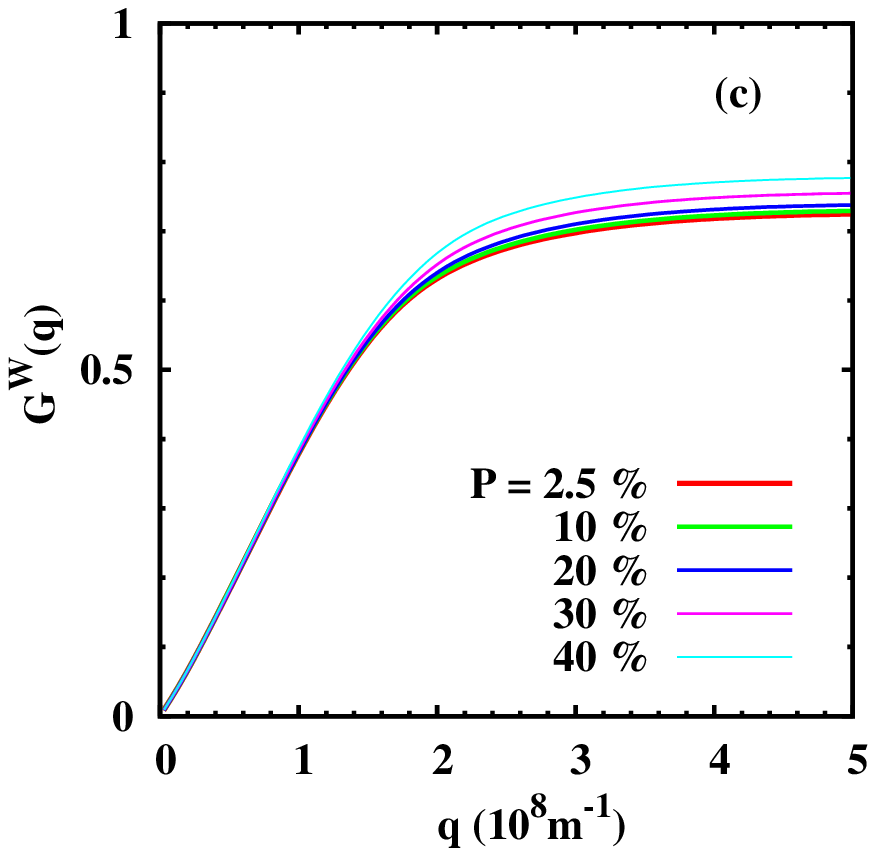}
\caption{(Color online) The local field factor $G^{\mbox{\tiny W}}(q)$ versus the
  wave vector in GaAs/Al$_{0.4}$Ga$_{0.6}$As QWs
with well width $a=10$\ nm. (a) At different temperatures
when electron density $n = 2.5 n_0$; (b) For different electron densities at $T=30$\ K;
  (c) For different spin polarizations at $T=30$\ K and $n = 2.5 n_0$. }
\label{figG}
\end{figure}

We plot the wave-vector dependence of
$G^{\mbox{\tiny W}}(q)$ in GaAs/Al$_{0.4}$Ga$_{0.6}$As QWs
with well width $a=10$\ nm under different conditions in Fig.\ \ref{figG}.
In Fig.\ \ref{figG}(a),  $G^{\mbox{\tiny W}}(q)$ are plotted at different
temperatures ($T$) when the electron density $n = 2.5 n_0 = 2.5 \times 10^{11}$\ cm$^{-2}$ (the corresponding
  $r_s=1.09$). One can see that
$G^{\mbox{\tiny W}}(q)$ increases very quickly
with temperature. The electron-density dependence of
 $G^{\mbox{\tiny W}}(q)$ is plotted in Fig.\ \ref{figG}(b) at
$T=30$\ K, where the largest electron density is chosen to be $5 n_0$  ($r_s=0.77$), still beyond the RPA
limit. It is seen that
$G^{\mbox{\tiny W}}(q)$ decreases with electron density $n$.
We further present  $G^{\mbox{\tiny W}}(q)$ at different spin
polarizations $P$ when $n = 2.5 n_0$ and
$T=30$\ K. One finds that $G^{\mbox{\tiny W}}(q)$ increases slightly with $P$.
When we use the  infinite-well-depth assumption and
let the well width to be very small, our $G^{\mbox{\tiny W}}(q)$ tends to
$G_{\mbox{\tiny 2D}}(q)$.\cite{bulutay}

Facilitated with the local field factor,
we turn to investigate the effect of
LFC on spin dephasing in $n$-type (001) GaAs/Al$_{0.4}$Ga$_{0.6}$As QWs.
We construct the kinetic spin Bloch equations by using the
  nonequilibrium Green function method:\cite{wu6}
  \begin{equation}
\dot{\rho}_{{\bf k},\sigma \sigma^{\prime}}=\dot{\rho}_{{\bf k},
\sigma \sigma^{\prime}}|_{coh}
+\dot{\rho}_{{\bf k},\sigma \sigma^{\prime}}|_{scatt},
\label{bloch}
\end{equation}
where $\rho_{{\bf k},\sigma \sigma^{\prime}}$ denote the
single particle density matrix elements with the diagonal and off-diagonal
elements to be the electron distribution functions $f_{{\bf k}\sigma}$ and spin
coherence $\rho_{{\bf k},\sigma-\sigma}$.
The coherent terms $\dot{\rho}_{k,\sigma \sigma^{\prime}}|_{coh}$ describe the precession
of the electron spin due to the effective magnetic field
$\mathbf{\Omega}({\bf k})$ [Eqs.\ (\ref{omegax}-\ref{omegaz})] as well
as the HF Coulomb interaction.\cite{wu3}
In the scattering term $\dot{\rho}_{k,\sigma \sigma^{\prime}}|_{scatt}$,
the electron-LO-phonon, the electron-AC-phonon,
the electron-nonmagnetic impurity and the electron-electron Coulomb
scattering are included explicitly. Their expressions can be found
in Refs.\ [\onlinecite{wu1,wu4}].
After numerically solving the kinetic spin Bloch equations, one can
obtain the spin dephasing and relaxation times from the
temporal evolutions of the spin coherence $\rho_{{\bf k},\sigma-\sigma}$
and the electron distribution function $f_{{\bf k},\sigma}$.\cite{wu7}

\begin{figure}[htb]
\includegraphics[height=5.5cm]{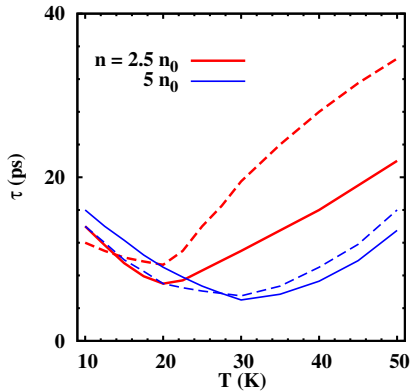}
\caption{(Color online) SRT versus temperature with (solid curves) and without (dashed
  curves) the LFC for  two different electron densities $n = 2.5 n_0$ and $5n_0$. $a=10$\ nm and
  $P=2$\ \%.}
\label{fig2}
\end{figure}

In Fig.\ \ref{fig2}, we present the temperature dependence of the spin
relaxation time (SRT) in QW with ($\tau_{\mbox{\tiny LFC}}$) and without
($\tau$) the LFC for two different electron
densities $n = 2.5 n_0$ and $5 n_0$.
The spin polarization is small ($P=2$~\%) in the calculation. The well
 width $a=10$\ nm. We choose the impurity density to be $0.0084 n_{0}$,
corresponding to that we used in Ref.\ \onlinecite{wu8}.
In our calculation, we focus on low temperature case.
It is seen from the figure that
$\tau_{\mbox{\tiny LFC}}$ is a little larger than $\tau$ at very low
temperature and becomes smaller than $\tau$ at higher temperatures for
both cases. These results are due to the reduction of the Coulomb scattering.
It has been pointed out
that the SRT decreases with the total scattering strength ($1/\tau_{p}^\ast$ with
$\tau_{p}^\ast$ representing the momentum relaxation time\cite{wu1,harley}) in the
weak scattering limit ($|\mathbf{\Omega}|\tau_{p}^\ast > 1$), but increases with the total
scattering strength in the
strong scattering limit ($|\mathbf{\Omega}|\tau_{p}^\ast \ll 1$).\cite{wu5,wu8}
Therefore, at very low temperature, the system is in the
weak scattering limit and the decrease of the Coulomb scattering leads to
$\tau_{\mbox{\tiny LFC}} > \tau$.
At higher temperature, however, the system is in the strong scattering limit.
Consequently, with the decrease of the Coulomb scattering, $\tau_{\mbox{\tiny LFC}} < \tau$.
It should be noted that the amplitude of the modification of the SRT
is no more than one order of magnitude because of
the existence of other  scattering terms such as the electron-phonon and the electron-impurity
scattering.

\begin{figure}[htb]
\includegraphics[height=5.5cm]{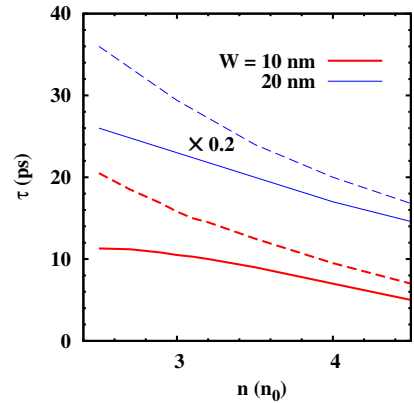}
\caption{(Color online) SRT versus electron density with (solid curves) and without (dashed
  curves) the LFC for two different well widths  $a=10$\ nm and
$20$\ nm.  $T=30$\ K and $P=2$ \%. The SRTs
  for the case with $a=20$\ nm are re-scaled by multiplying 0.2.}
\label{fig3}
\end{figure}

From Fig.\ \ref{fig2}, one also finds that the reduction of the SRT by the LFC
for the case with $n = 2.5 n_0$ is  much larger than that with $n =
5 n_0$. Therefore, it is necessary to further investigate 
 the density dependence of the SRT with/without the LFC. In Fig.\ \ref{fig3}
the SRTs are plotted against the electron density $n$
 for different well widths $a=10$\ nm and
$20$\ nm when  $T= 30$\ K and $P = 2$ \%. In order to make the figure clear, we
re-scale the SRTs for the case with $a=20$\
 nm by multiplying 0.2. As the system is in the strong scattering limit, $\tau_{\mbox{\tiny LFC}} < \tau$.
Moreover, the larger the electron density is, the smaller the difference between
$\tau_{\mbox{\tiny LFC}}$ and $\tau$ becomes. This is because
the reduction of the Coulomb scattering due to
the LFC becomes weaker for larger electron density. This can be seen
from  Fig.\ \ref{figG}(b) that
 $G(q)$ decreases with $n$, especially at large $q$ near the Fermi
 momentum $k_F$.

\begin{figure}[htb]
\includegraphics[height=5.5cm]{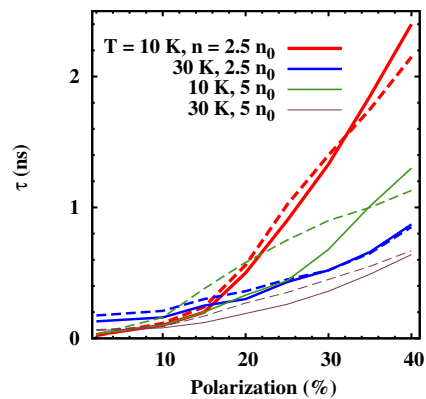}
\caption{(Color online) SRT versus spin polarization with (solid curves) and without (dashed
  curves)  the LFC for two temperatures $T=10$\ K and $30$\ K and two electron densities $n = 2.5 n_0$ and $5 n_0$.
 $a=20$\ nm.}
\label{fig4}
\end{figure}

Finally, we investigate the spin polarization dependence of $\tau$ and
$\tau_{\mbox{\tiny LFC}}$ for different temperatures and electron densities as
shown in Fig.\ \ref{fig4}. In the calculation, the well width  $a=20$\ nm. It is interesting to see that
there is a crossover at certain spin polarization.
$\tau_{\mbox{\tiny LFC}} < \tau$ for small spin polarization and $\tau_{\mbox{\tiny LFC}} > \tau$ for large
spin polarization. These features origin from the dual effects of the Coulomb interaction at
high spin polarization. The first effect  has been said above.
The second one is from the effective magnetic field in the HF term which
suppresses the spin R/D at high spin polarization.\cite{wu3}
 This effect has been observed in experiments very recently.\cite{wu8,zheng}
As the LFC always reduces the Coulomb interaction,
the LFC can lead to a decrease
of the SRT by reducing the HF term at high spin polarization.
It is seen from the figure that for small spin polarization, as the contribution
from the HF term is negligible and the system is in the strong
scattering limit, inclusion of the LFC always results in a decrease of the SRT.
However, when the polarization becomes large enough, the contribution from the HF term
becomes important and the system turns into the weak scattering limit as the HF term enhances the amplitude
  of effective magnetic field $|\mathbf{\Omega}|$ along the $z$-axis.\cite{wu3,wu8}
In this regime, inclusion  of the LFC results in a competing effects from the
HF term and the scattering term. Our results indicate that when the polarization is
large enough, the effect from the scattering is stronger and consequently the SRT
is increased with the inclusion of the LFC.

In conclusion, we study the STLS LFC on
the spin relaxation/dephasing in $n$-type GaAs (001) QWs at low temperature by
  constructing and numerically solving the kinetic spin Bloch
equations. The LFC takes into account of the screening from
the additional exchange-correlation hole ignored in the RPA.
The LFC reduces both the Coulomb scattering and the Coulomb HF interaction.
It is known that the decrease of the Coulomb scattering
can lead to a decrease (an increase) of
the SRT in the strong (weak) scattering limit
whereas the decrease of the  Coulomb HF interaction
results in a decrease of the SRT at high spin polarization.\cite{wu3}
After comparing the SRT with and without the LFC
under different conditions, we find that the LFC can lead to a
decrease  of the SRT in the strong  scattering limit for small spin polarization. It also leads to a
decrease of the SRT  while the spin polarization getting larger, and finally leads to an increase when the spin
 polarization is large enough.

It is noted that in our recent theoretical
comparison with  the experiment,
the LFC was ignored.\cite{wu8} From this investigation, we conclude that this
approximation is acceptable.
This is because when the electron density is about $2.1 n_{0}$,
the difference between $\tau_{\mbox{\tiny LFC}}$ and $\tau$
is very small according to Fig.\ \ref{fig4}.

The author would like to thank M. W.Wu for proposing the topic as well as the directions
during the investigation.
This work was supported by the Natural
Science Foundation of China under Grant Nos.\ 10574120 and
10725417, the
National Basic Research Program of China under Grant
No.\ 2006CB922005 and the Knowledge Innovation Project of Chinese Academy
of Sciences.


\begin{thebibliography}{0}
\bibitem{comment} It is noted that in most litereature, it says that the RPA is good only
when $r_s\ll1$. However, there is also claim that the RAP is good
upto $r_s=5$ [see, S. Adam, E. H. Hwang, V. M. Galitski, and S. Das Sarma, PNAS {\bf 104},
18392 (2007)].
\end{thebibliography}
\end{document}